# Species-Dependent Electron Emission from Nanoparticles under Gamma Irradiation

Darukesha B H M

*Indian Space Research Organization, Bengaluru 560 017, India*
*Faculty of Space Technologies, AGH University of Krakow, 30-065, Poland*

## Abstract

In this study, various nanoparticle species—including Au and $Gd_2O_3$—were irradiated with low-energy gamma rays, such as 59.5 keV photons from $^{241}Am$. Pulse-height spectra were recorded using a liquid-scintillation counting system before and after dispersing the nanoparticles into the scintillator, and the differences between them were analyzed to infer the interaction outcomes. $Gd_2O_3$ nanoparticles emitted numerous electrons; however, under identical experimental conditions, no detectable electron emission was observed from Au nanoparticles (AuNPs). Here, "detectable electron emission" refers to electrons with energies high enough to be registered by the liquid-scintillation detector used ($\approx$ 100 eV and, more typically, $\geq$ 1–2 keV); however, it excludes electrons that may be emitted at lower energies. Thus, a species-dependent radiation–nanoparticle interaction was observed. Rigorous controls and falsification tests excluded various artefacts—including detector insensitivity, surface contamination, aggregation, quenching, and self-absorption—as causes for the absence of detectable electron emission from AuNPs. This observation potentially prompts a re-evaluation of the common assumption that nanoparticles behave like their bulk counterparts emit electrons upon gamma-irradiation. Instead, our results suggest that the distinct internal environment of nanoparticles influences their interaction with radiation. These findings offer significant insights for practical applications, including a better mechanistic understanding of nanoparticle radiosensitization in cancer therapy, enhanced gamma-detection efficiency of organic scintillators, and the development of lightweight radiation shields.

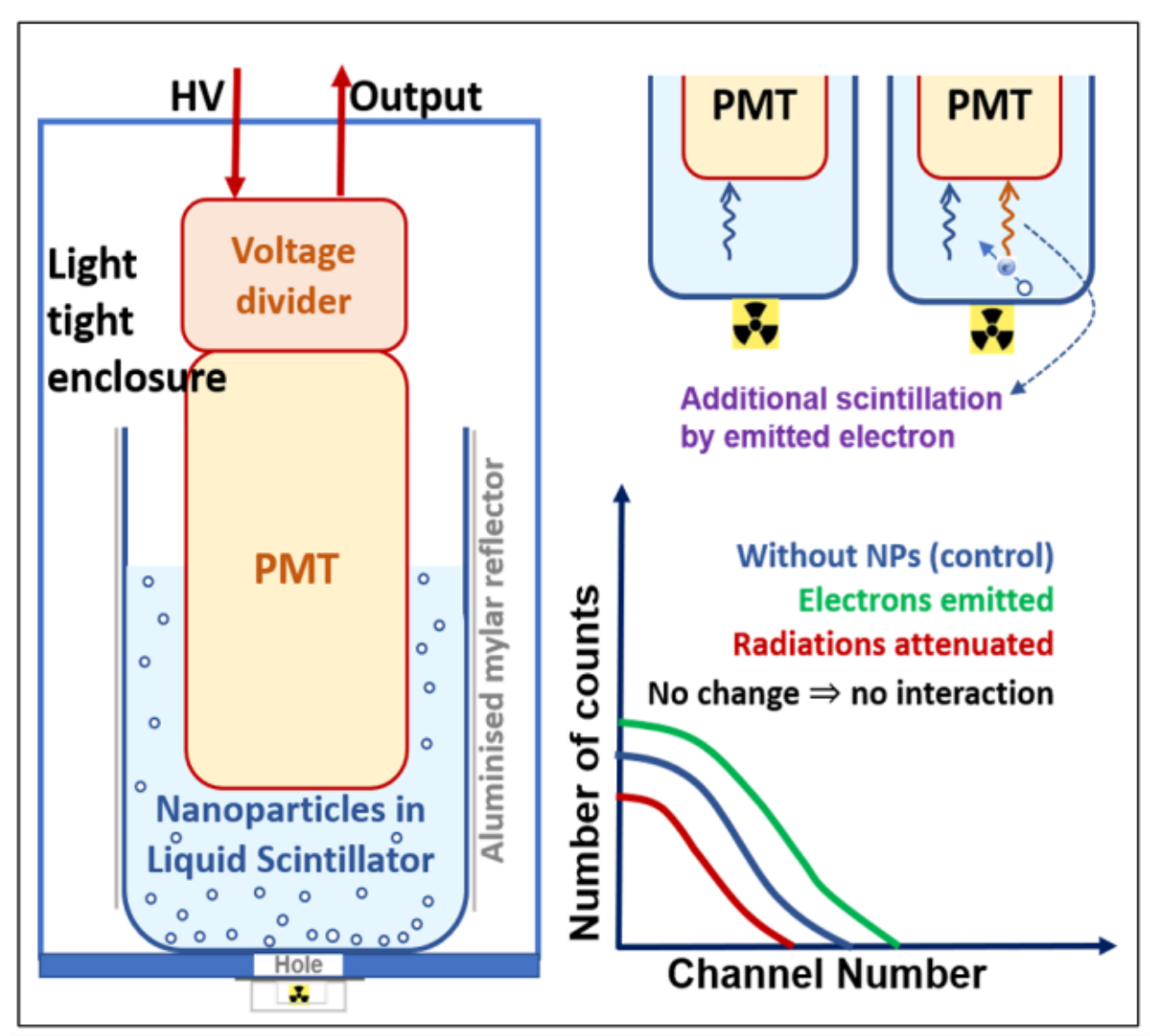

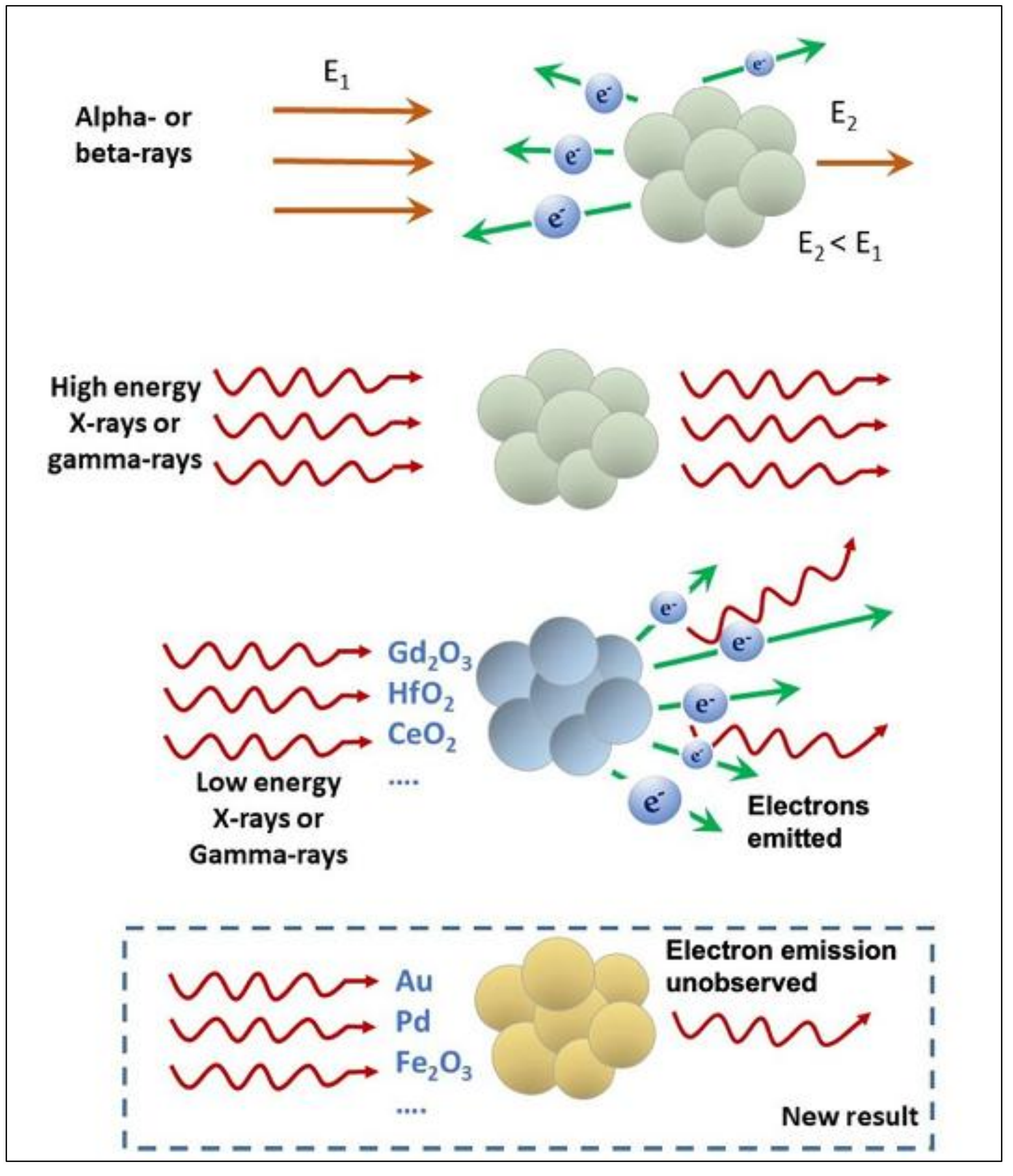

**Graphical representation of the method and the new result**

## 1. Introduction:

Nanoparticles have emerged as versatile agents in radiation science, with applications ranging from radiosensitization in cancer therapy to enhancing gamma-detection efficiency of organic scintillators and the development of lightweight gamma-ray shields (B. H. M. Darukesha et al., 2023a; Liu et al., 2024; Naik et al., 2024). Central to many of these applications is the emission of electrons from nanoparticles upon exposure to ionizing radiation, which can contribute to nanoparticle radiosensitization (NPRS) or improve detector efficiency and shielding effectiveness. Prevailing interaction models often assume that nanoparticles exhibit electron emission properties analogous to those of their bulk material counterparts (Casta et al., 2015; Hajagos et al., 2018; Liu et al., 2024; Schuemann et al., 2020). However, this assumption lacks robust empirical support, a gap acknowledged by the research community (Brun & Sicard-Roselli, 2016; Coughlin et al., 2020; Korman et al., 2021).

In our previous work, we incorporated nanoparticles of $Gd_2O_3$, $HfO_2$, $Bi_2O_3$, $WO_3$, and PbO into separate plastic scintillators to compare the efficacy of nanoparticles in enhancing the gamma-ray detection efficiency of the host materials (B. H. M. Darukesha et al., 2023b). The rationale is that the host material is an inefficient gamma-ray detector. However, in nanoparticle-loaded scintillators, gamma rays interacting with nanoparticles are expected to emit photoelectrons into the host matrix. These electrons would then interact with the scintillator, increasing scintillation counts, a concept explored in the literature (Koshimizu, 2020; McKigney et al., 2007). However, we observed that $Gd_2O_3$, $HfO_2$, and $Bi_2O_3$ nanoparticles could enhance the detection efficiency, whereas those loaded with $WO_3$ and PbO could not, leading us to infer that these nanoparticles may inhibit electron emission into the host material. To the best of our knowledge, enhancement of gamma-detection efficiency in plastic scintillators by incorporating $WO_3$ or PbO nanoparticles is yet to be reported in the literature, despite the extensive research on loading these nanoparticles in polymers for gamma-shielding.

The above observations motivated us to develop a simple, more direct method to determine electron emission from nanoparticles upon gamma-ray irradiation. In response, we employed a liquid scintillation counting system for this purpose. We recorded pulse height spectra with a given radiation source, before and after dispersing the nanoparticles into the scintillator under identical measurement conditions, and compared the spectra. Differences between the two spectra revealed the nature of the interactions: an increase in counts indicates detectable electron emission from the nanoparticles, as emitted electrons interact with the scintillator medium and generate additional scintillation events; a reduction in counts implies attenuation of the incident gamma rays by the nanoparticles; and an unaltered spectrum suggests that gamma rays pass through the nanoparticles without significant interaction.

In this study, using the technique described above, we investigated electron emission from a range of nanoparticle species dispersed in the liquid scintillator system and irradiated with low-energy photons. Here, we refer to X-rays from $^{55}Fe$ or a 40 kVp gun, and gamma rays from $^{241}Am$ or $^{133}Ba$ as low-energy photons. The study reveals notable differences in electron emission behavior among nanoparticle species: $Bi_2O_3$, $CeO_2$, $Eu_2O_3$, $Gd_2O_3$, HfO2, $LaF_3$, $SnO_2$, and $TiO_2$ nanoparticles emit electrons, whereas Au, CuO, $Fe_2O_3$, PbO, Pd, $Pr_2O_3$, Sn, W, $WO_3$, and Zn nanoparticles show no detectable electron emission, with the term 'detectable' denoted in the abstract.

For the nanoparticle species and low-energy photon interactions investigated, these findings necessitate a re-examination of the assumption that all nanoparticle species emit electrons into their surroundings. They potentially impact ongoing practical applications as they provide important mechanistic insights into NPRS, a means to choose suitable nanoparticles as hosts for enhancing gamma-detection efficiency of scintillators and for developing lightweight radiation shields. It appears that the internal environment of nanoparticles may differ from that of bulk materials in ways that decisively influence the nature and outcome of these interactions.

## 2. Experimental

### Materials & Methods

While comprehensive details of our experimental setup and procedures are available in the thesis (Darukesha, 2023), this section provides all essential information required to replicate the core experiments. Most nanoparticles used in this study were commercially sourced, while gold nanoparticles (AuNPs) were both sourced and synthesized in-house. The sourced nanoparticles, along with their suppliers, purities, average sizes, and polydispersity indices (PDI), are listed in Table 1.

Table 1 Nanoparticles used in the study

| NP | Source | Purity (%) | Average Size (nm) | PDI |
|---|---|---|---|---|
| Au | Nanoshel | 99.9 | 73.0 ± 3.3 | 0.316 |
| $CeO_2$ | Otto Chemie | 99.9 | 38.5 ± 1.7 | 0.287 |
| CuO | Otto Chemie | 99.9 | 112.4 ± 2.3 | 0.360 |
| $Eu_2O_3$ | Mincometsal | 99.9 | 74.9 ± 1.5 | 0.333 |
| $Fe_2O_3$ | High Purity Lab Chemie | 99.4 | 32.9 ± 0.1 | 0.250 |
| $Gd_2O_3$ | Nanoshel | 99.9 | 55.0 ± 0.8 | 0.292 |
| $LaF_3$ | Lab Chemie | 99.7 | 41.9 ± 2.3 | 0.298 |
| PbO | Nanoshel | 99.9 | 141.0 ± 2.4 | 0.299 |
| Pd | Lab Chemie | 99.8 | 33.1 ± 1.1 | 0.254 |
| $Pr_2O_3$ | Mincometsal | 99.9 | 116.9 ± 2.1 | 0.364 |
| Sn | Otto Chemie | 99.7 | 10.5 ± 8.9 | 0.297 |
| $SnO_2$ | High Purity Lab Chemie | 99.9 | 83.8 ± 1.2 | 0.301 |
| $TiO_2$ | Sigma-Aldrich | 98.7 | 36.3 ± 2.9 | 0.262 |
| W | High Purity Lab Chemie | 99.8 | 70.7 ± 1.6 | 0.320 |
| $WO_3$ | Nanoshel | 99.9 | 81.6 ± 1.4 | 0.351 |
| Zn | Otto Chemie | 99.7 | 128.0 ± 5.1 | 0.308 |
| ZnO | Nanoshel | 99.9 | 73.0 ± 3.3 | 0.316 |

Additionally, citrate-buffered AuNPs with core sizes of 4–7 nm and mean hydrodynamic diameters of 14–25 nm were sourced from Sigma-Aldrich.

### Synthesis of Citrate-Buffered AuNPs

Citrate-buffered AuNPs were synthesized in-house using the Turkevich method (Turkevich et al., 1951). High-resolution transmission electron microscopy images of the synthesized AuNPs, obtained with an FEI-Titan Themis 300 kV microscope, are shown in figure 1. The average particle size was 29 ± 0.3 nm with a polydispersity index of 0.352.

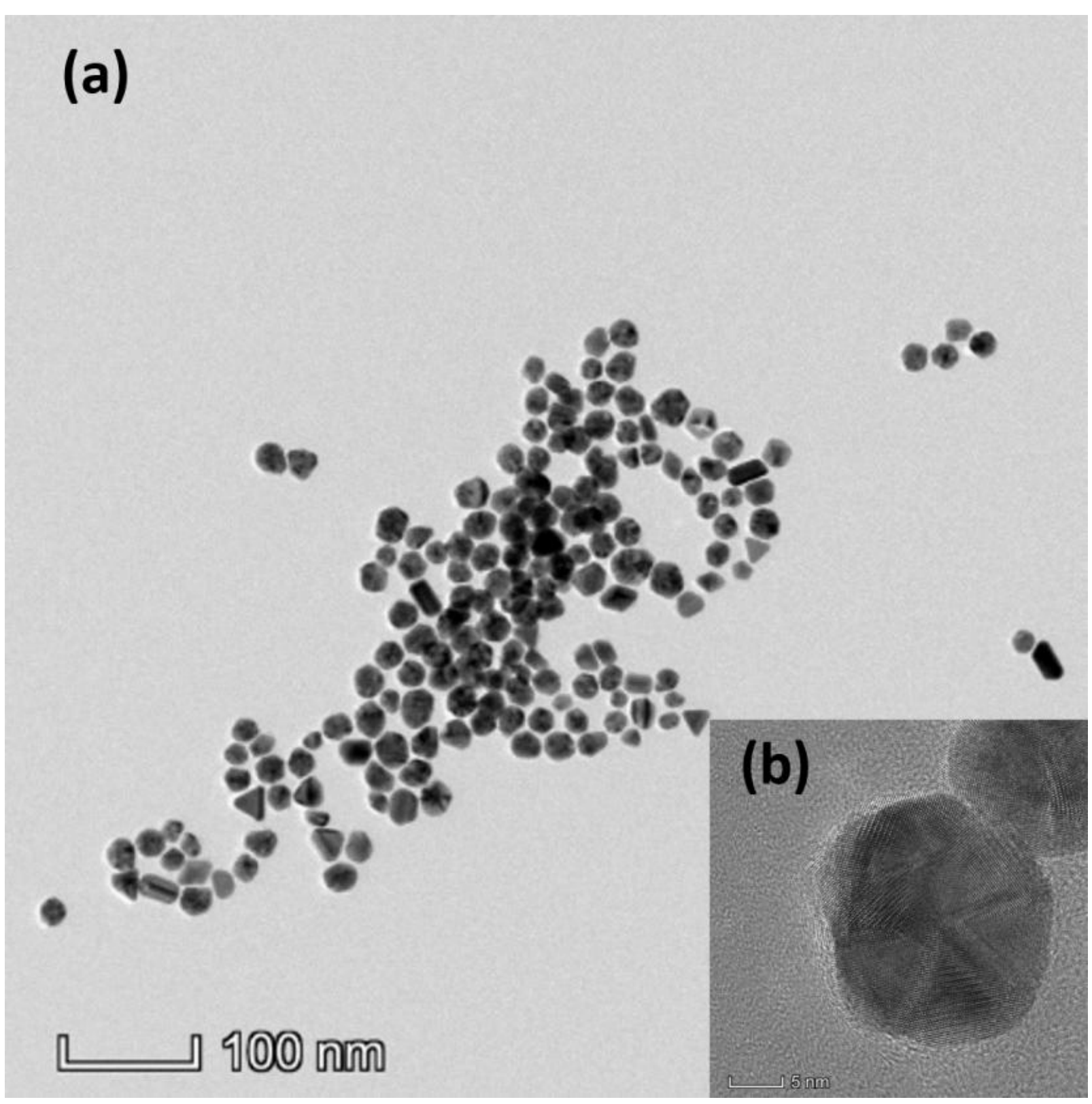

**Fig. 1**. **(a)** Transmission electron micrograph of AuNPs synthesized **(b)** A high-resolution micrograph of the synthesized AuNPs

A conventional toluene-based liquid scintillator was sourced from SRL Chemicals (srlchem.com, Product #46069). It consists of 6 g of 2,5-diphenyloxazole (PPO) fluorophore and 0.2 g of 1,4-bis(5-phenyloxazol-2-yl) benzene (POPOP) wavelength shifter dissolved in 1 liter of toluene. Additionally, a dioxane-based liquid scintillator was obtained from Mayochem (mayochem.com, Product #23072). This scintillator contains 1,4-dioxane as the solvent, with 120 g/L naphthalene, 4 g/L PPO, and 50 mg/L POPOP.

Gamma ray sources used in this study were $^{241}$Am and $^{133}$Ba, while $^{90}$Sr/$^{90}$Y served as the beta radiation source. All three sources were in the form of discs, approximately 3 mm thick and 10 mm in diameter, supplied by BRIT (britatom.gov.in). A Moxtek Magnum 40 kVp X-ray gun was also employed, emitting approximately $3 \times 10^{11}$ photons/sr at 100 µA; it was operated at 38 kVp and 50 µA throughout the experiments. Additionally, a button source of alpha-emitting $^{241}$Am was obtained from an outdated smoke detector.

Au:Sn (80:20) solder alloy sheets, 25 µm thick, were sourced from Indium Corporation of America.

Nanoparticle characterization techniques employed in this study included Scanning Electron Microscopy with Energy Dispersive Spectroscopy, Dynamic Light Scattering (DLS), X-ray Photoelectron Spectroscopy (XPS), X-Ray Diffraction, Spectrophotometry, Transmission Electron Microscopy, and Photoluminescence, as detailed in our previous work (B. H. M. Darukesha et al., 2023a) and thesis (H. M. Darukesha, 2023)

The method used to determine the interaction outcomes of gamma rays, X-rays, and beta radiation is outlined in the Introduction. For alpha radiation measurements, the $^{241}$Am alpha button source was immersed directly into the liquid scintillator, as alpha particles cannot penetrate the thickness of the glass beaker. Because the $^{241}$Am source also emits gamma rays, the detector's high voltage (HV) was reduced from the typical 800 V used in other tests to 600 V. This reduced voltage was sufficient to register interactions from alpha particles but low enough to suppress signals from gamma rays emitted by the same source. Consequently, at 600 V, the detected signal arises exclusively from alpha particle interactions with the nanoparticles.

**Uncertainty and Error Analysis**

The study compares PHS obtained under identical conditions before and after loading nanoparticles into the scintillator. Background count rates in the PHS were as low as 0.7, 3.8, and 7.8 counts per second (cps) at HVs of 600 V, 800 V, and 1000 V, respectively. Background count rates measured with sources positioned under empty beakers were indistinguishable from those obtained with the radioactive sources alone. Background counts were not subtracted from the plotted data as they were negligible and consistent across measurements. Positioning the X-ray gun without a scintillator yielded a count rate of 35 cps, which increased to 14,850 cps when the beaker contained liquid scintillator. The HV remained constant during acquisition, ensuring a stable detector response.

Repeatability was confirmed through at least 12 independent measurements performed over several weeks. Variations in total count rates remained within approximately 1%, expressed as the standard deviation (SD) of these measurements. For example, in a 59.5 keV gamma-ray test, the unloaded scintillator yielded 28,183 ± 310 cps (mean ± SD, n = 12), while the addition of 50 mg $Gd_2O_3$ increased the count rate to 116,933 ± 719 cps (mean ± SD, n = 12). Count reductions relative to the unloaded scintillator were consistently reproducible across independent repeats and nanoparticle batches, and dilution studies confirmed signal stability. Nanoparticles were weighed using a Sartorius Genius scale with a resolution of 10 µg for sub-milligram quantities.

To characterize the detection threshold of the scintillator, a 4 kVp X-ray gun—the lowest energy photon source available in our setup—was used. X-ray guns generate photons through Bremsstrahlung, producing a continuous energy spectrum ranging approximately from near zero up to the peak kilovoltage (kVp) of the tube, with a weighted average photon energy typically about one-third of the peak kVp. Hence, we state that the detection threshold of the present scintillator is typically in the range of 1–2 keV.

**Rationale for the electron emission determination method**

The interpretation of pulse height spectrum (PHS) changes relies on the fundamental operation of the liquid scintillation detector. Energy deposited in the scintillator excites the matrix, producing UV photons that are rapidly shifted into the visible range by embedded wavelength shifters. Photomultiplier tubes (PMTs) then convert this visible light into an electronic pulse, giving direct access to the underlying radiation–matter interaction. A PHS is simply a histogram that counts how many detector pulses occur at each signal amplitude, so it maps the distribution of energies deposited by the incoming radiation.

In the unloaded scintillator, atoms such as C, H, N, O in solvents like toluene or dioxane interact with incoming gamma rays, absorbing a fraction of the radiation energy. The dissolved fluorescent materials (fluorophore and wavelength shifter) become excited upon receiving this energy, and their subsequent de-excitation releases scintillation light. PMT kept in optical contact with the liquid scintillator detects this light and converts it into fast electrons. These electrons are then multiplied by the high-voltage grids within the PMT, producing a current pulse at the anode. This current is converted into a voltage pulse and registered as a count in the PHS. A scintillation event (triggered by an electron interaction) generates a pulse, and the amplitude of this pulse (related to scintillation intensity) reflects the energy deposited. The readout system sorts electrons by their deposited energy, displaying this distribution as the PHS characteristic of the unloaded scintillator.

When nanoparticles are dispersed into the scintillator, they also interact with the incoming gamma rays. Portions of the gamma-ray energy are converted into electrons within the nanoparticles. If and when emitted, these electrons interact with the surrounding scintillator medium. While gamma-ray interactions are stochastic, most emitted electrons will *likely* deposit energy in the scintillator, although very low-energy electrons may be absorbed before producing detectable light. The scintillator converts all fast-electron tracks to photons, independent of the electron's origin. Therefore, an increase in the number of electrons emitted by nanoparticles increases counts in the PHS, providing a measure of electron emission.

## 3. Results and inference

Figure 2 shows the variations in the PHS of gamma rays from $^{241}$Am for the unloaded liquid scintillator upon incremental loading of gold nanoparticles (AuNPs). The X-axis shows channel numbers proportional to deposited energy. We observed a reduction in counts across all channels upon nanoparticle loading, with further reductions corresponding to increased loading. Both buffered and unbuffered AuNPs led to a decrease in counts, as shown in Figure 3.

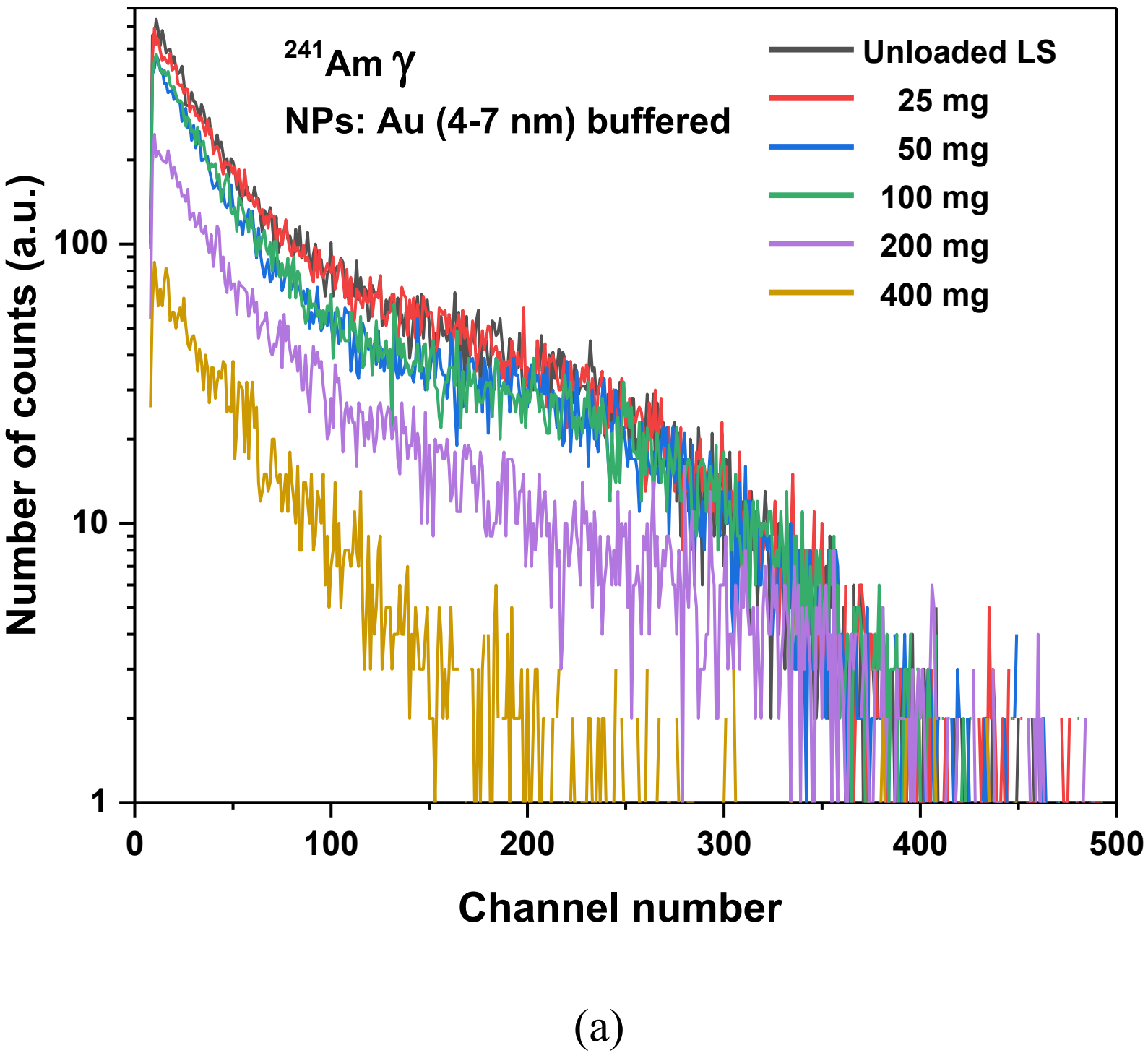

(a)

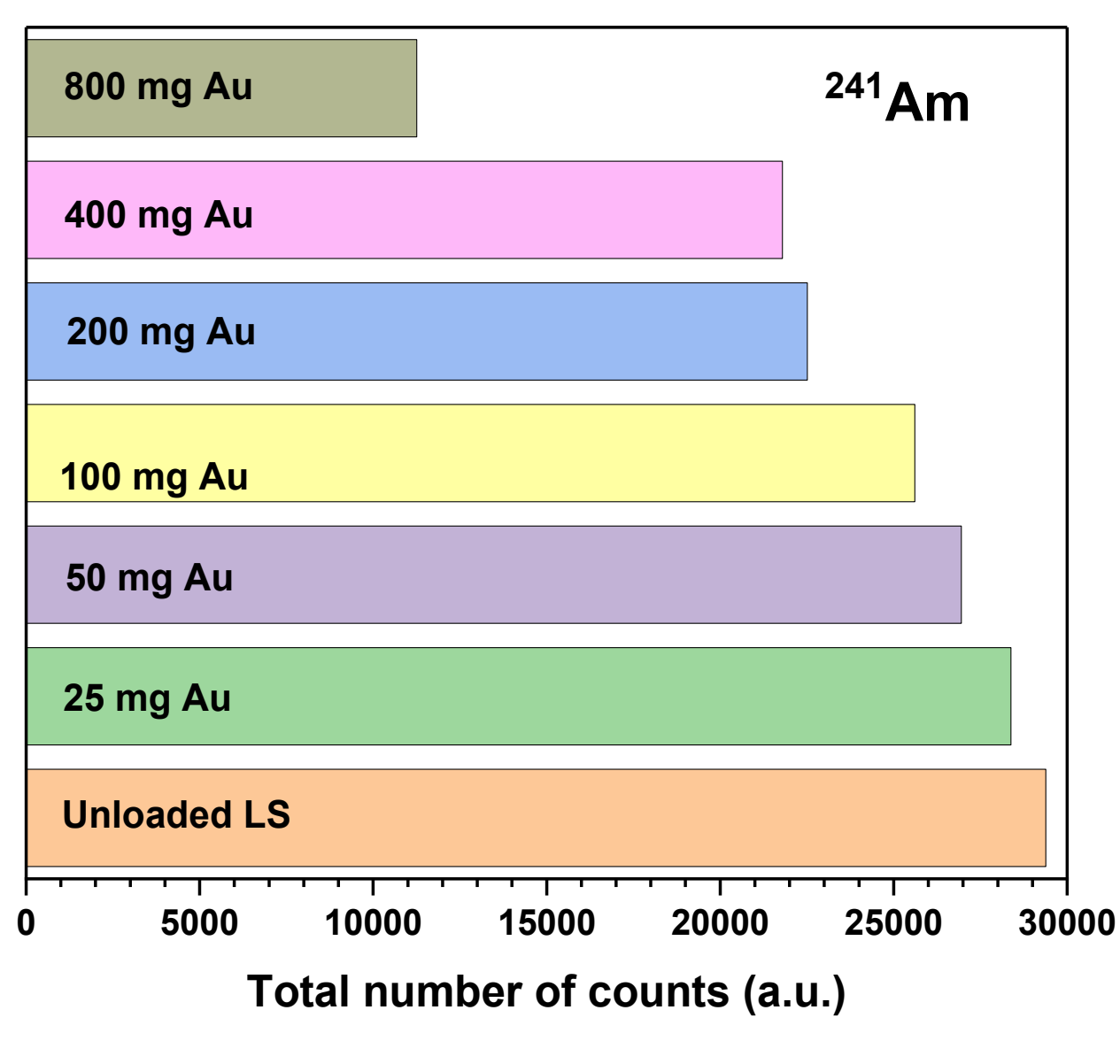

(b)

**Fig. 2. (a)** Variations in the pulse height spectrum of gamma rays from $^{241}$Am for the unloaded liquid scintillator upon loading gold nanoparticles. **(b)** Histograms showing the total counts in the PHS upon loading AuNPs and Pd nanoparticles

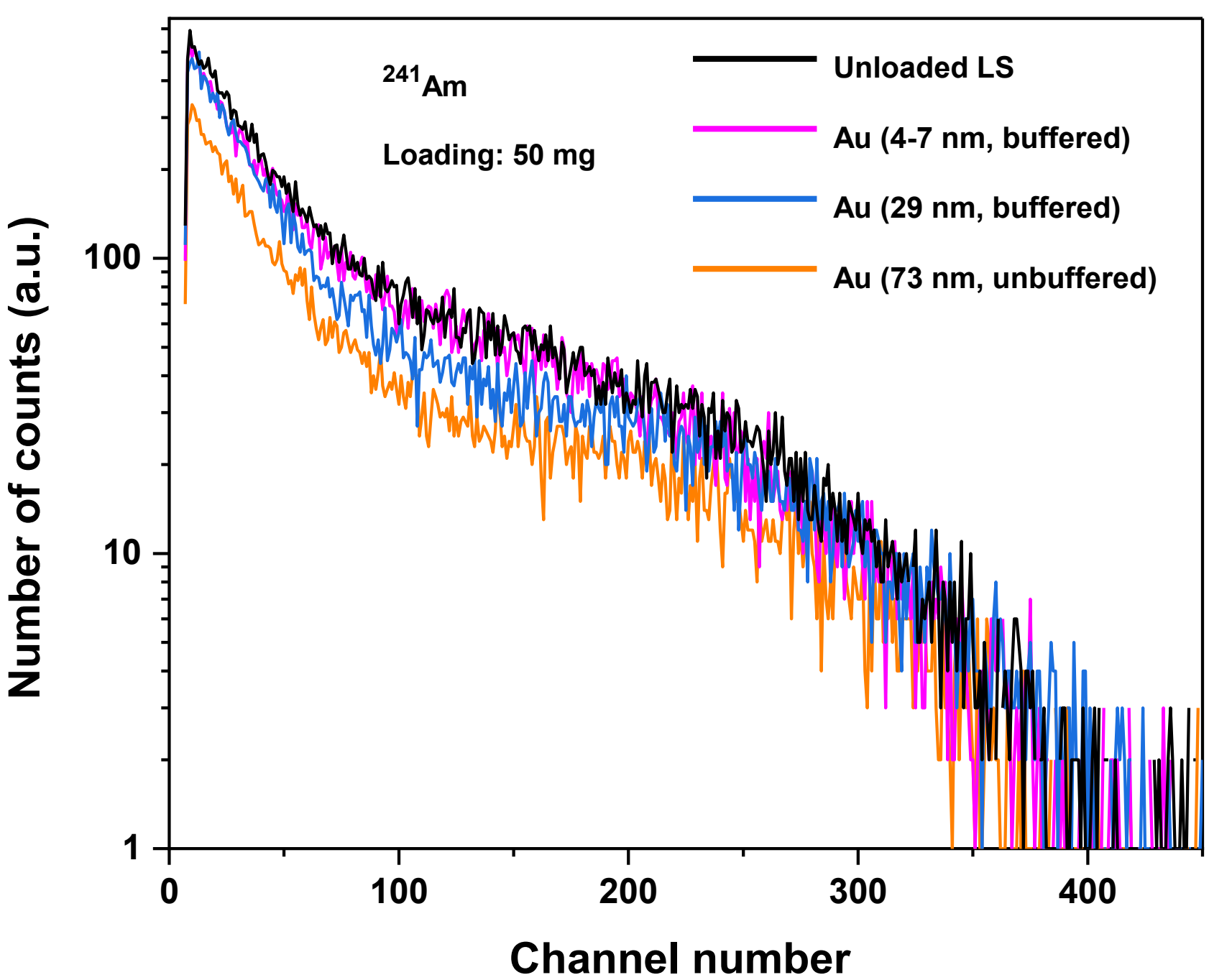

**Fig. 3.** Variations in the pulse height spectrum of gamma-rays from $^{241}$Am for the liquid scintillator upon loading gold nanoparticles of different sizes

Based on the rationale of the technique, the observed reduction in counts suggests that AuNPs effectively attenuate photons within themselves, and the electrons produced are not detectably emitted into the surrounding scintillator medium.

Further, the interaction outcomes between low-energy photons—X-rays from a 40 kVp source and $^{55}$Fe, as well as gamma rays from $^{133}$Ba—and AuNPs or $Gd_2O_3$ nanoparticles followed trends similar to those for $^{241}$Am gamma rays. Nanoparticles such as Au, CuO, $Fe_2O_3$, PbO, Pd, $Pr_2O_3$, Sn, W, $WO_3$, and Zn exhibited interaction profiles comparable to AuNPs, characterized by a reduction in counts indicating attenuation effects. In contrast, nanoparticles including $Bi_2O_3$, $CeO_2$, $Eu_2O_3$, $Gd_2O_3$, HfO2, $LaF_3$, $SnO_2$, and $TiO_2$ displayed responses similar to those of $Gd_2O_3$ nanoparticles, associated with increased counts indicating detectable electron emission. Experiments conducted using both a dioxane-based liquid scintillator and Hidex's Aqualight+ Ultra Low Level™ liquid scintillator yielded consistent outcomes.

To explore a broader gamma-ray energy range, measurements were conducted using photons from $^{137}$Cs, $^{60}$Co, and $^{22}$Na sources. Interactions of these high-energy photons with AuNPs resulted in negligible changes in the PHS at low nanoparticle concentrations, with a reduction in counts observed only at higher loadings. This pattern was consistent across all nanoparticle species tested. Therefore, further investigations using high-energy gamma rays were discontinued.

The observed reduction in counts upon loading AuNPs into the liquid scintillator and irradiating with low-energy photons was surprising. In contrast to the negligible variations seen with high-energy gamma rays, the reduction in PHS counts confirms significant photon–nanoparticle interactions at low energies. Under identical conditions, $Gd_2O_3$ nanoparticles produced an increase in counts of up to approximately 3900% (B. H. M. Darukesha et al., 2023a), highlighting that these effects are intrinsic to the nanoparticles rather than arising from experimental artifacts. Nevertheless, a thorough investigation and analysis were undertaken to rule out artifacts and potential alternative explanations systematically.

**Investigation and analysis**

Liquid scintillation counters (LSCs) are a well-established and extensively researched technique for detecting beta radiation used for over six decades. The development of the radioimmunoassay technique, which relies on LSC for measuring radioactivity in biological samples led to thorough investigation of the technique. In 1970, Aschorft loaded Sn atoms into toluene-based LSCs and demonstrated enhanced gamma ray detection capabilities (Ashcroft, 1970). LSCs remain widely used in fields such as environmental monitoring (Li et al., 2022), radioactivity measurement of low-energy isotopes, pharmaceutical research (Schubert & Kallmeyer, 2023), and neutrino detection (Zhang et al., 2024). In this study, nanoparticles were dispersed directly in the liquid scintillator—an inherently efficient electron detector—to enable the detection of electron emission from the nanoparticles themselves.

Determination and comparison of the shape of the PHS when the gamma rays are attenuated outside the scintillator by a known material would validate the setup. It was investigated by placing a 25 μm-thick Au:Sn (80:20) alloy sheet between the gamma ray source and the unloaded liquid scintillator. The resulting PHS is shown in Figure 4. It closely resembles that obtained with AuNPs dispersed inside the scintillator. It shows a similar reduction in counts to that obtained with AuNPs dispersed inside the scintillator. This similarity suggests that, like the alloy sheet, the AuNPs attenuate a portion of the gamma rays, either while suspended in the scintillator or after settling at the bottom. A microsyringe was employed to load buffered AuNPs into the liquid scintillator. Notably, a 25 μL volume of buffered AuNPs (4–7 nm)—the minimum volume drawable with the microsyringe—corresponds to an estimated layer thickness of approximately 67 μm in the beaker. In both cases, the number of gamma rays reaching the scintillator is reduced.

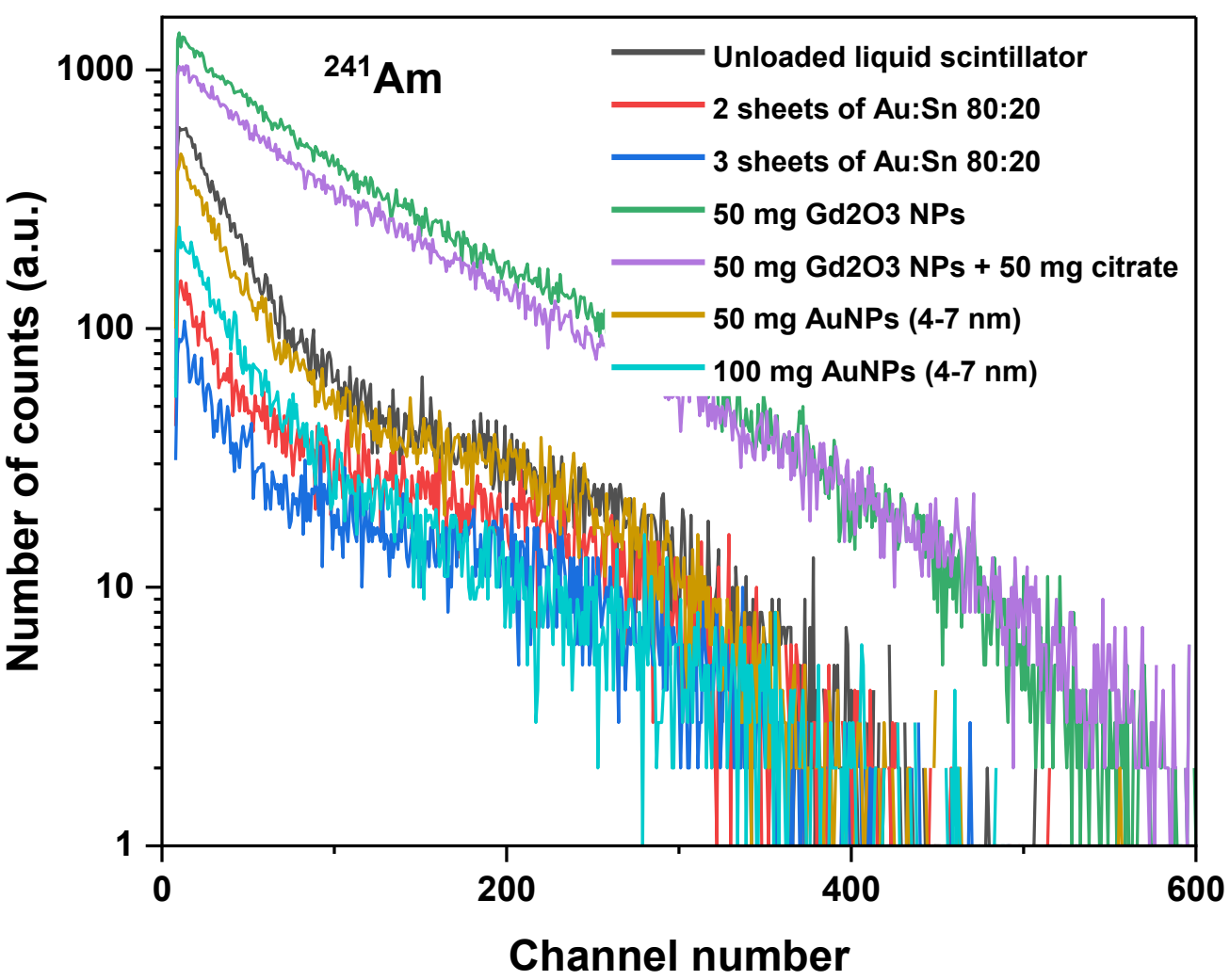

**Fig. 4.** The variations in PHS of gamma rays from $^{241}$Am for unloaded liquid scintillator upon inserting Au:Sn 80:20 sheets between the source and the beaker. Also displayed are the variations in PHS for unloaded liquid scintillator upon loading citrate solution.

Citrate coating on buffered AuNPs may absorb some of the emitted electron energy (Beeler et al., 2017). To investigate whether electrons exiting AuNPs are absorbed by the citrate layer, PHS variations were recorded under three conditions: (i) loading citrate buffer alone, (ii) loading buffered AuNPs, and (iii) loading unbuffered $Gd_2O_3$ nanoparticles together with neat citrate buffer (Figure 4). The addition of 50 mg of neat citrate buffer to the liquid scintillator caused a minor reduction in counts of approximately 1%; this difference is not shown in the plot as it is indistinguishable. Loading 50 mg of buffered 4–7 nm AuNPs reduced counts by about 9% (Figures 2b and 4). In contrast, loading 50 mg of unbuffered $Gd_2O_3$ nanoparticles together with 50 mg of neat citrate increased counts by approximately 330%, reaching 93,000 counts. These results suggest that the citrate buffer may not substantially absorb electron energy emitted by the nanoparticles.

Nanoparticles tend to aggregate over time. Experiments with the same batch of $Gd_2O_3$ nanoparticles were conducted over a year to assess consistency. A batch of AuNPs synthesized in-house was first tested for their gamma ray response on the day of synthesis, before characterization by TEM. Certain nanoparticles were sintered to obtain larger counterparts: 55 nm $Gd_2O_3$ nanoparticles were sintered to average particle sizes of 179 nm and 1249 nm; 81 nm $WO_3$ nanoparticles were sintered to 880 nm; and 73 nm ZnO nanoparticles to 630 nm. $Gd_2O_3$ nanoparticles of all tested sizes and 73 nm ZnO nanoparticles emitted electrons, while both 81 nm $WO_3$ nanoparticles and their larger sintered counterparts attenuated photons. Observations with the 630 nm ZnO nanoparticles are discussed later. These demonstrations, together with those including under charged-particle irradiation, rule out aggregation—whether aggregation occurred before or after dispersion—as a factor influencing electron emission.

Contamination on the surface of nanoparticles can potentially hinder electron emission. To rule out this possibility, nanoparticles were irradiated with beta and alpha radiation. The pulse height spectra for beta radiation from $^{90}Sr/^{90}Y$ and alpha particles from $^{241}Am$ showed an increase in counts upon loading nanoparticles into the liquid scintillator (Figure 5), consistent across all species. Counts increased at lower loading but decreased at higher concentrations. The liquid scintillator itself is an effective beta detector; when beta radiation strikes nanoparticles, they can generate additional electrons at the entry point, within the core, and at the exit—provided sufficient energy remains to escape. These escaping beta radiation and secondary electrons further interact with the scintillator and nearby nanoparticles. At higher loadings, densely packed nanoparticles form a layer that attenuates incoming alpha or beta radiation. These results confirm that electrons generated by beta and alpha radiation do exit the nanoparticles. While beta and alpha interactions differ fundamentally from gamma interactions, this finding suggests that gross surface contamination preventing electron escape is unlikely.

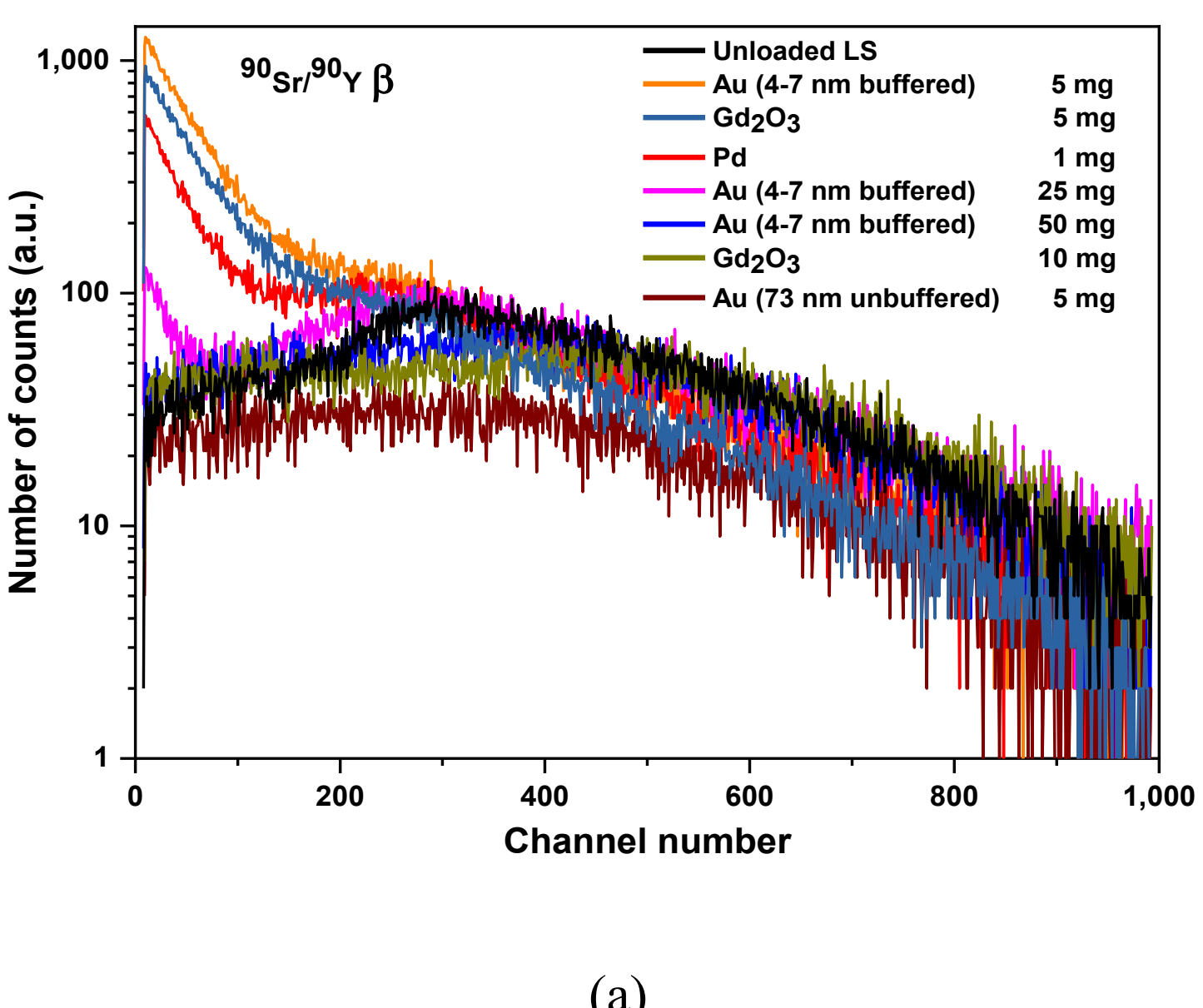

(a)

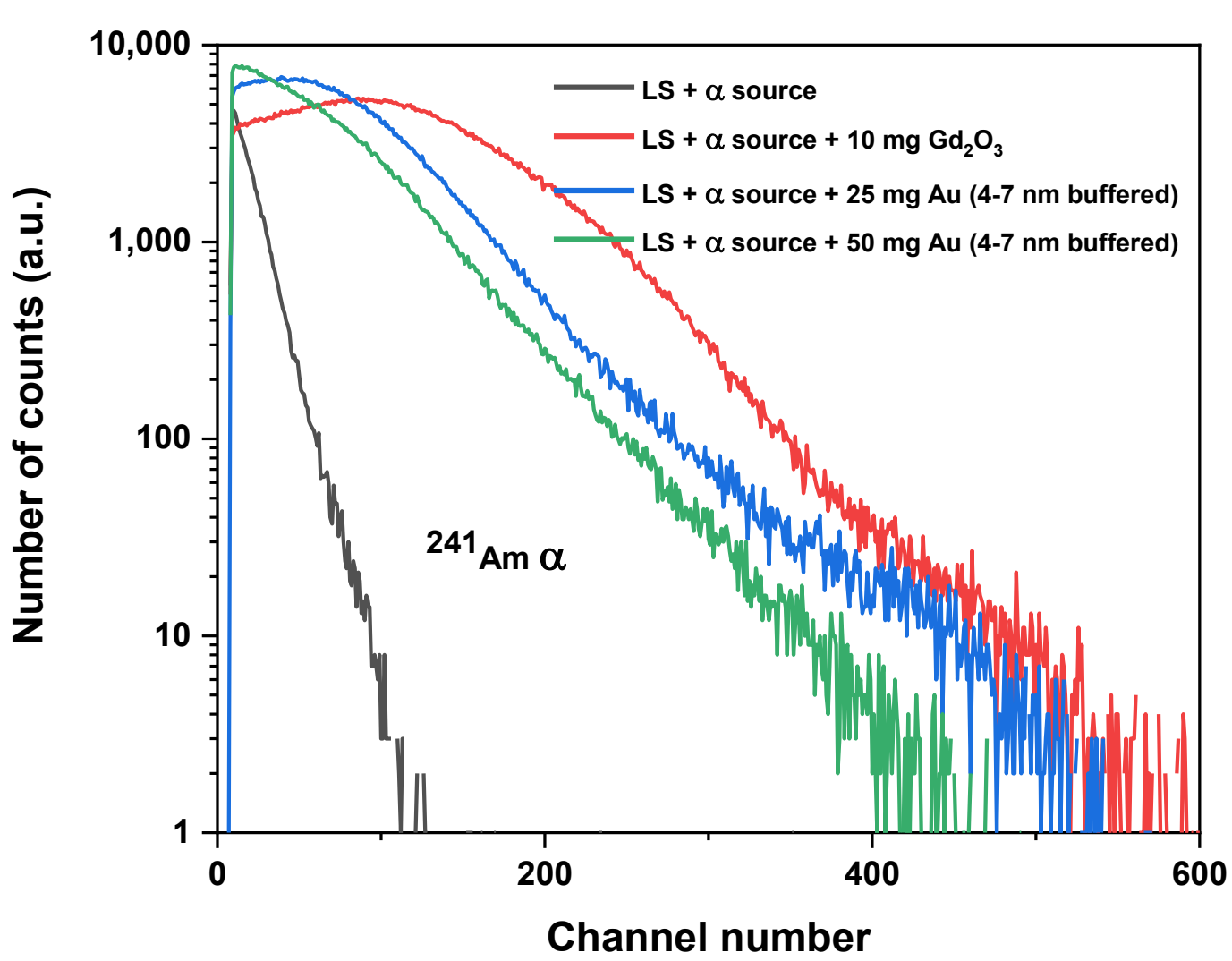

(b)

**Fig. 5.** The variations in PHS of a beta radiation from **(a)** $^{90}Sr/^{90}Y$ and **(b)** alpha-radiation from $^{241}Am$ upon loading the nanoparticles

The physical distribution of nanoparticles in the liquid scintillator—whether suspended or sedimented—does not significantly affect the results. Dispersion characteristics in the present solution are comparable to those observed in DLS with toluene. Toluene was used in DLS to determine the sizes of these nanoparticles. Hence, the dispersion status of nanoparticles is guaranteed for several minutes. At typical loadings, a large portion of nanoparticles stay suspended in the liquid scintillator for tens of minutes to hours before settling at the bottom, while the measurement time is typically 2 minutes after loading the nanoparticles. The reduction in counts from PHS obtained for the first two minutes of loading 25 mg of $Gd_2O_3$ nanoparticles to that after 90 minutes of loading in an undisturbed setup was less than 2%. Even in a settled configuration, sedimentation can provide a favorable condition by creating a uniform nanoparticle layer that interacts effectively with the radiation flux, provided the loading or sediment thickness is not sufficient to absorb radiation or emitted electrons. This also helps clear the scintillator above of potential optical quenching effects, owing to effective source–detector geometry. Dilution studies and charged-particle radiation experiments (Figure 5) confirm that sedimentation is not responsible for the lack of detectable electron emission observed under gamma-ray irradiation.

It is possible that the initial nanoparticle loadings were too high to allow emitted electrons to effectively interact with the liquid scintillator. As established in our previous study, all nanoparticle species begin to attenuate gamma rays once loaded beyond a species-dependent threshold, effectively behaving like a thick layer. To address this, loadings that caused reduced counts—such as those with AuNPs—were lowered, and experiments were repeated at these reduced concentrations. Although progressively lower loadings caused the PHS to gradually approach the baseline, they never exceeded the baseline (control) PHS, and no increased counts were observed at any reduced loading.

The minimum volume of buffered AuNPs drawable with a microsyringe was as low as 25 µL. Figure 6 illustrates the effect of decreasing loadings on the PHS, exemplified by Pd nanoparticles. When loading exceeded 0.1 mg, the mixture was diluted by shaking, decanting approximately half (~3 mL), and replenishing with fresh liquid scintillator. Despite dilution, counts remained reduced relative to the unloaded scintillator. Thus, for certain nanoparticle species, detectable electron emission was not observed at any tested loading, down to the lowest practical loading.

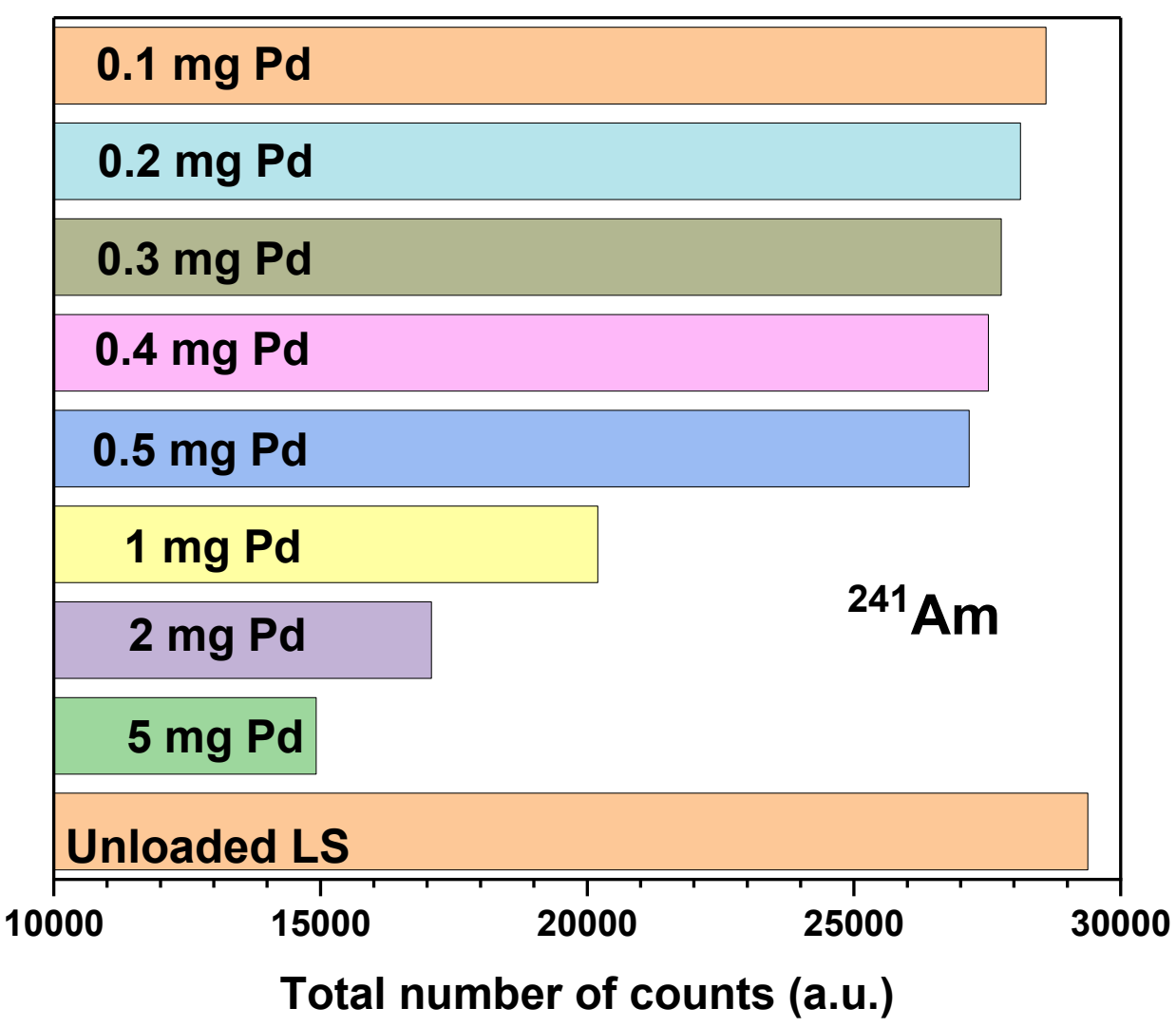

**Fig. 6.** The variations in PHS of gamma-rays from $^{241}$Am for liquid scintillator upon loading Pd nanoparticles

Optical quenching could potentially explain the observed reduction in PHS counts. Optical quenching refers to the decrease in fluorescence emission intensity of a fluorophore due to nonradiative energy transfer or absorption by nearby quenchers, such as metal nanoparticles. Since nanoparticles exhibit some degree of quenching, if quenching were the primary factor, the significant enhancement of counts observed upon loading $Gd_2O_3$ nanoparticles would be unlikely. Moreover, as shown in Figure 5, counts increased substantially under beta and alpha irradiation for the same nanoparticle–liquid scintillator combinations, effectively ruling out optical quenching as the main cause during gamma ray interactions. Buffered AuNPs are denser than the scintillator and tend to settle at the bottom, resulting in a sparse suspension of nanoparticles. Once electrons or scintillation photons exit the nanoparticle layer, the likelihood of optical quenching is minimal.

The observed reduction in counts cannot be attributed to self-absorption of scintillation light by the nanoparticles. The PPO fluorophore in the scintillator emits light primarily between 350–400 nm, peaking at 360 nm, while the POPOP wavelength shifter absorbs this emission and re-emits light between 380–450 nm, peaking at 420 nm. AuNPs exhibit strong absorption around 512 nm, which is well outside the emission range of both PPO and POPOP. Similarly, $WO_3$ and PbO nanoparticles do not absorb photons at these wavelengths (B. H. M. Darukesha et al., 2023b). Therefore, self-absorption effects by the nanoparticles are unlikely to cause the observed reduction in PHS counts. Thus, other mechanisms must be responsible for the count reductions observed in the experiments.

The nanoparticles in the present study were exposed to a high-intensity gamma-ray flux. Based on the activity, live time, and solid angle, they receive more than $10^5$ gamma rays per second from $^{241}$Am and approximately $10^8$ X-rays per second with a weighted average energy of about 10 keV. The schematics and geometry are the same as provided in previous work (B. H. M.

Darukesha et al., 2023a). Although the estimated probability of interaction between gamma rays (at energies typically used in NPRS) and AuNPs is quite low—ranging from approximately $10^{-5}$ to $10^{-4}$ (Lin et al., 2014; Schuemann et al., 2020)—Even if only 1% of the nanoparticles convert gamma rays into electrons, about 25% of ionizations result in the emission of one or two electrons from AuNPs (McMahon et al., 2011). Therefore, the calculated flux appears adequate for detecting electron yields predicted by standard cross-section models, provided the electrons exceed the detector threshold.

A liquid scintillator system has a scintillation threshold—the minimum energy required to produce detectable light. For toluene-based scintillators, this threshold is estimated to be approximately 100 eV (Perera et al., 1992). Precise characterization for a specific system remains challenging (Kessler & Kessler, 2015). In our experiments with low-energy photon sources such as $^{55}$Fe and a 4 kVp X-ray gun, the system produced clear PHS responses, as shown in Figure 7. The formation of distinct PHS peaks confirms that the setup is sensitive to electrons with energies down to approximately 100 eV, consistent with the weighted average energy of 1-2 keV from the X-ray gun. A precise determination of the scintillation threshold would tighten the lower-energy limit for electrons that our detector can register. Accordingly, for each of the photon energies listed above, we recorded how the count rate changed after adding $Gd_2O_3$ nanoparticles and, separately, gold nanoparticles (AuNPs) to the scintillator.

The HV applied to the PMT influences the minimum electron energy that can be accelerated and detected. The high voltage was set as high as 1000 V in this ET Enterprises 9078B PMT. However, the results remained unchanged across this entire voltage range. The experiments were conducted with nanoparticles $Gd_2O_3$ having a mean diameter of 12.96 nm and a PDI of 0.428; the findings were consistent with those obtained for $Gd_2O_3$ nanoparticles of other sizes. However, due to the unavailability of ultrasmall AuNPs (<4–7 nm), this study did not investigate whether such particles exhibit different electron emission behaviors. Notably, these limitations do not invalidate the findings on the species-dependent nature of electron emission.

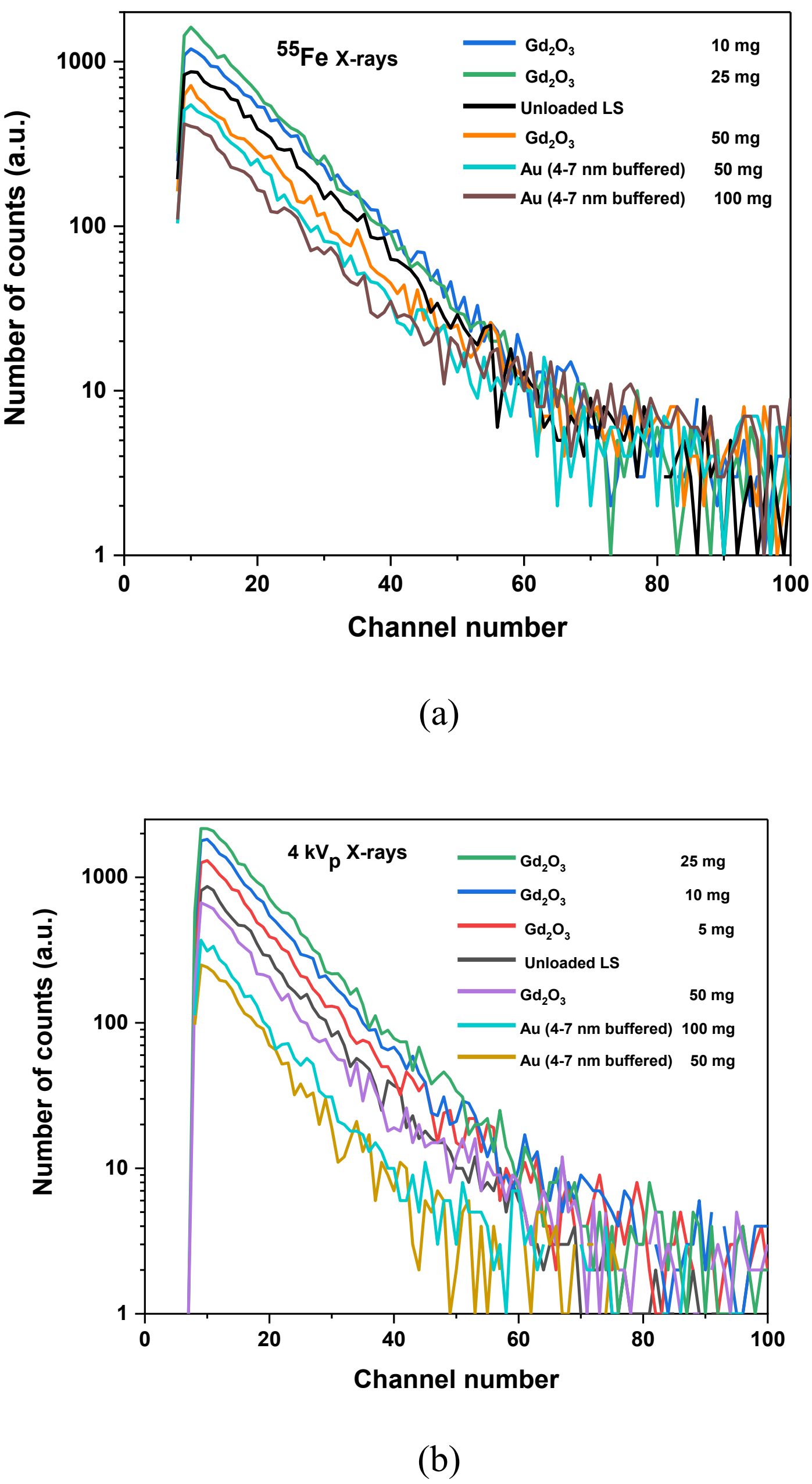

(a)

(b)

**Fig. 7**. The variation in PHS for unloaded liquid scintillator for interaction with X-rays from **(a)** $^{55}Fe$ and **(b)** an X-ray gun at 4 $kV_p$ upon loading $Gd_2O_3$ and AuNPs

Analytically, photoelectrons emitted from gold (Au) upon interaction with gamma rays from $^{241}Am$ and $^{133}Ba$ possess energies up to approximately 46 keV and 67 keV, respectively. The estimated continuously slowing down approximation (CSDA) ranges—the average distances that charged particles such as electrons travel in a material before coming to rest—for electrons of these energies in gold exceed 4000 nm and 6000 nm, respectively (physics.nist.gov). Because the CSDA range exceeds the nanoparticle diameter by one to two orders of magnitude, *most* photo- and Auger electrons are expected to exit the particle. Auger electrons emitted from Au upon interaction with 59.5 keV photons are expected to have energies between 1 and 10 keV, with many near ~1 keV. In comparison, Fe- and Cu-based nanoparticles exposed to $^{241}Am$, the X-ray gun, or $^{133}Ba$ sources are expected to emit Auger electrons with characteristic energies around 5.5 keV and 6.9 keV, respectively.

Thus, through a series of control experiments, known artifacts and potential alternative explanations for the observed signals have been ruled out. Analytical considerations of electron energies and ranges further support the expectation that photoelectrons generated within nanoparticles should escape and be detectable under the experimental conditions. Despite this, the absence of detectable electron emission from certain nanoparticle species suggests that these particles primarily attenuate low-energy photons rather than emitting electrons. While this study has limitations, the analyses presented offer insights that merit further investigation

## 4. Discussion

In the context of existing literature, the present study provides initial experimental evidence suggesting that nanoparticles can emit electrons when irradiated with photons of defined energies. The findings endorse those obtained through an indirect method in our previous study, where $Gd_2O_3$, $HfO_2$, and $Bi_2O_3$ enhanced the gamma-detection efficiency of plastic scintillators, while $WO_3$, PbO failed to do so (B. H. M. Darukesha et al., 2023b). Indeed, there has been a notable lack of empirical techniques capable of directly probing how photons at energies relevant to practical applications interact with nanoparticles (Brun & Sicard-Roselli, 2016; Coughlin et al., 2020; Schuemann et al., 2020).

Materials characterization techniques such as X-ray photoelectron spectroscopy (XPS) involve irradiating a material with X-rays and detecting the emitted photoelectrons. However, XPS typically uses photon energies up to approximately 1.5 keV and detects electrons emitted into a vacuum rather than within a medium. In a comparable setup, (Casta et al., 2014) reported that the number of photoelectrons emitted from gold nanoparticles (AuNPs) was only marginally higher than that from bulk gold surfaces, attributing this difference to nanoparticle size effects.

Nanoparticle radiosensitization (NPRS) is an approach designed to enhance the efficacy of radiotherapy by administering nanoparticles into malignant tissues before irradiation. Most NPRS studies infer outcomes from a complex interplay of physical, chemical, and biological processes, making it challenging to isolate the contribution of each component. In contrast, the present study specifically focuses on the initial physical interaction phase between radiation and nanoparticles, aiming to clarify this fundamental step.

While the NPRS technique is well-established, its underlying mechanistic basis remains debated due to surprising findings, and this lack of clarity has hindered further development and rapid clinical implementation (Brun & Sicard-Roselli, 2016; Butterworth et al., 2012; Casta et al., 2015; Coughlin et al., 2020; Loscertales et al., 2025; Luo, 2025; McMahon et al., 2011; Schuemann et al., 2020). The prevailing hypothesis of NPRS posits that nanoparticles interact more readily with incoming radiation than tumor tissue, emitting photoelectrons or Auger electrons that induce chemical and biological effects to eradicate cancer cells or inhibit their growth (Brun & Sicard-Roselli, 2016; Butterworth et al., 2012; McMahon et al., 2011; Schuemann et al., 2020).

Gold nanoparticles (AuNPs) are the most extensively studied species and are favored due to their high atomic number, synthetic controllability, and biocompatibility (Butterworth et al., 2012; Coughlin et al., 2020; Gilles et al., 2014; Schuemann et al., 2020). However, definitive experimental evidence of detectable electron emission from AuNPs under low-energy photon irradiation remains limited (Coughlin et al., 2020). Notably, smaller AuNPs used as contrast agents attenuate photons more effectively than larger counterparts or iodine-based agents (Chenjie et al., 2008). Radiosensitization with nanoparticles has been demonstrated under high-energy gamma irradiation (MeV range), despite the low interaction probability at these energies (Brun & Sicard-Roselli, 2016; Schuemann et al., 2020). Meanwhile, nanoparticles such as $HfO_2$ and $Gd_2O_3$ are advancing more rapidly in preclinical radiosensitization trials (Brun & Sicard-Roselli, 2016; Jayakumar et al., 2017; Schuemann et al., 2020). For its mechanistic understanding, several other factors are hypothesized to cause NPRS.

The emission of low-energy electrons (LEEs) has been proposed as an alternative mechanism for NPRS (Sanche, 2016), and the present study cannot exclude their involvement if their energies fall below the scintillation threshold of the liquid scintillator used. Notably, the experimental setup closely simulates real-world field conditions, enhancing the relevance and applicability of the findings. The liquid scintillator employed here has an effective atomic number and density comparable to water. Even if emitted, electrons do not appear to cause detectable excitation within this scintillator. While the scintillator has properties comparable to water, further dedicated studies are needed to confirm if a similar lack of detectable excitation would occur in water. Under the conditions of this study, photoelectron or Auger electron emission from certain nanoparticles was not detected above the system's sensitivity threshold.

Collectively, the results of the present study and observations in the literature suggest that NPRS may arise through mechanisms beyond the *detectable* direct photoelectron emission, with higher-energy electron emission hypothetically playing a supportive rather than a dominant role. However, this study cannot exclude the involvement of very low-energy electrons (LEEs) if their energies fall below the scintillation threshold, which could still represent a dominant contribution to NPRS.

Nanoparticles are embedded in plastic scintillators to enhance the gamma-detection efficiency of the host, as discussed in the Introduction. It has been demonstrated with the incorporation of $ZrO_2$, $LaF_3$, $Gd_2O_3$, but, as noted in the Introduction, such enhancement is yet to be reported in the literature with $WO_3$, Sn, PbO nanoparticles. The nanoparticles in the latter set could have been a logical first choice, since the loading of Pb and Sn atoms—not the nanoparticles—into liquid and plastic scintillators enhances the gamma-detection efficiency of the hosts, as demonstrated and utilized since the 1960s (Ashcroft, 1970). Meanwhile, these nanoparticles are incorporated into polymers to develop efficient, lightweight, flexible gamma-ray shields, since the existing shields—including those worn by radiation personnel—are heavy and less flexible. While any species of nanoparticles is suitable, there are reports of $WO_3$ and Pb-based nanoparticles offering remarkably high shielding efficiency. The literature attributes the remarkable gamma shielding ability of the nanocomposites to the exceptionally high packing density of the nanoparticles (Ambika et al., 2017; Noor Azman, 2013). Also, it is desirable to use nanoparticles that undergo interactions resulting in electron emission, particularly low-energy electrons. Otherwise, the thickness and thus the weight of the shield would increase to stop those energetic electrons.
Thus, the empirical findings from the present technique, complemented by existing literature, suggest reconsideration of the assumption of bulk-like response of nanoparticles in radiation interactions. Nevertheless, before adopting these implications in practical applications, it is necessary to verify electron emission using alternative techniques. Scintillation detectors are a standard and reliable method for detecting electrons. One potential approach is to irradiate these nanoparticles inside an XPS system using a $^{241}$Am source. Another promising method involves using channel electron multipliers to amplify the emitted electrons and assess whether similar results can be obtained. Additionally, irradiating nanoparticles with these sources inside particle detectors presents another viable possibility for further verification.

**Interpretation and Outstanding Questions**

We are actively investigating the mechanisms responsible for the lack of detectable electron emission observed in specific nanoparticle species. Fundamentally, the initial photon interactions (e.g., photoelectric absorption) with individual atoms may not differ significantly between bulk materials and their nanoparticulate counterparts composed of the same element. However, our findings suggest that the subsequent electron transport and emission processes are critically modified by the distinct internal environment of nanoparticles, leading to species-dependent outcomes.

Notably, the gamma-detection efficiency of liquid scintillators was observed to improve when Pb and Sn atoms—not nanoparticles—were incorporated into them, as reported in studies from the 1970s (Ashcroft, 1970). However, similar enhancements have not been demonstrated upon loading their nanoparticle forms. Among the interaction mechanisms, the photoelectric effect remains dominant. Measurements using CdTe detectors further support this conclusion (B. H. M. Darukesha et al., 2023a). When the liquid scintillator was replaced with a CdTe detector, the counts decreased for all nanoparticle species, including $Gd_2O_3$, as expected, since CdTe detectors primarily detect X-rays rather than directly detecting electrons. Importantly, the reduction happened at photopeak.

Certain possibilities can, however, be ruled out when explaining the origin of these differences. Chemical composition alone does not account for the observed variations. For example, while W and $WO_3$ nanoparticles behave similarly to AuNPs, and Sn nanoparticles also exhibit AuNP-like behavior, $SnO_2$ emits electrons akin to $Gd_2O_3$. Similarly, magnetic properties do not fully explain the results: $Fe_2O_3$ and $Pr_2O_3$ respond similarly to AuNPs but differ from $Gd_2O_3$. Variations are also observed within transition metal nanoparticles. An intriguing, albeit speculative, observation is that nanoparticles exhibiting effective gamma ray attenuation in our study tend to be non-white in color. For instance, upon sintering, $Gd_2O_3$ nanoparticles retain their color and behave like $Gd_2O_3$, whereas ZnO nanoparticles darken and begin to behave like AuNPs. The fact that $Gd_2O_3$ nanoparticles sintered to an effective size of 2149 nm still emit electrons shows that the behavior of ZnO nanoparticles sintered to 630 nm cannot be explained by thickness effects alone. Also, color quenching has been excluded as a contributing factor. Nevertheless, this correlation requires dedicated investigation to establish any underlying physical basis.

Collectively, these observations suggest that, although photon interactions occur at the atomic scale, electron transport within nanoparticles plays a critical role in governing electron transmission and emission, leading to species-dependent outcomes.

## 5. Conclusions

This experimental study, focusing exclusively on the physical interactions between photons and nanoparticles, provides direct evidence that nanoparticles such as Au, Pd, $Fe_2O_3$, Sn, W, and $WO_3$ interact with low-energy photons yet produce no detectable electron emission above our ~ 100 eV and, more typically, ≥ 1–2 keV sensitivity threshold. These findings raise concerns about the prevailing assumption that all nanoparticle species interact with ionizing radiation similarly to their bulk counterparts. Although photon interactions occur at the atomic scale, electron transport within nanoparticles may be a significant factor influencing their ability to transmit and emit electrons upon irradiation. Further investigation is needed to confirm this hypothesis.

These findings provide crucial data that can help address a critical gap in the mechanistic understanding of nanoparticle radiosensitization (NPRS). The observed non-detection of electron emission from selected nanoparticle species, including AuNPs, upon interaction with low-energy photons, prompts a re-evaluation of the hypothesis that photoelectron or Auger electron emission constitutes the primary physical mechanism driving NPRS under these conditions, especially considering the acknowledged empirical gap in this area.

## Conflict of interest

There are no conflicts of interest.

## Acknowledgements:

BHMD gratefully acknowledges: Dr Radhakrishna V (Indian Space Research Organisation, ISRO), Prof. K Rajanna (IISc), Prof. Abha Misra (IISc), Dr M. Ravindra (RNSIT), Dr M. M. Nayak (CeNSE, IISc) for their guidance and support throughout the project; Dr Aruna S. T. (CSIR-NAL, Bengaluru) for supplying and sintering several of the nanoparticles used in this study; MNCF, IISc, Dr Parthasarathi Bera (CSIR-NAL), Dr S. M. M. Kennedy (SSNCE, Kalavakkam) and Dr Sarat Kumar Dash (ISRO) for nanoparticle characterisation; Shri Srikar Paavan Tadepalli and Shri Koushal Vadodariya for valuable discussions and instrumentation support; Smt M. A. Lalitha Abraham, Radiation Safety Officer, ISRO, for her vigilant oversight of radiation-safety protocols; the Directors, Heads, and staff of the Instrumentation Division and Space Astronomy Group at U R Rao Satellite Centre, Bengaluru.